\documentclass[aps,prl,twocolumn,showpacs,floatfix,preprintnumbers,amsmath,
amssymb,superscriptaddress]{revtex4}
\usepackage{graphicx}
\usepackage{color}

\newcommand{\beq}{\begin{equation}}
\newcommand{\eeq}{\end{equation}}
\newcommand{\beqn}{\begin{eqnarray}}
\newcommand{\eeqn}{\end{eqnarray}}
\newcommand{\bearr}{\begin{array}}
\newcommand{\enarr}{\end{array}}

\begin{document} 

\title{DNA denaturation and wetting in the presence of disorder}
      
\author{H. Kunz}
\affiliation{Ecole Polytechnique Federale de Lausanne,
SB-ITP-LPPC CH - 1015 Lausanne}
\author{R. Livi}
\affiliation{CSDC - Dipartimento di Fisica e Astronomia, 
Universit\`a di Firenze and INFN Sezione di Firenze,
via Sansone 1, I-50019 Sesto Fiorentino, Italy}

\begin{abstract}
We present a precise equivalence of the Lifson-Poland-Scheraga model with 
wetting models. Making use of a representation of the former model
in terms of random matrices, we obtain, in the limit of weak
disorder, a mean--field approximation, that shows a change of the
critical behavior due to disorder.
\end{abstract}

\pacs{05.70.Fh, 05.40.Fb, 02.50.-r, 87.15.-v}
\maketitle
Quenched disorder plays a crucial role in phase transitions. It can either
modify the nature or the order of the phase transition \cite{1,2,3}or
affect the value of the critical exponents \cite{4,5,6,7}.
In this letter we consider the Lifson--Poland--Scheraga
(LPS) model of DNA denaturation, in the limit of
weak quenched disorder.
In the ordered case, the order of the phase transition depends
crucially on a parameter $\alpha$ \cite{8}. It is continuous if
$1 < \alpha <2$, while the transition turns to first order if
$ \alpha > 2$. Recently, it has been rigorously proven \cite{9}
that the disordered model exhibits a continuos transition for
all values of $\alpha$. The values of the critical exponents
are unknown, except in the case $1 < \alpha < 3/2$, where they
coincide with the exponents of the ordered case \cite{10},
in agreement with Harris' criterion \cite{11}.
On the other hand, Giacomin and Toninelli \cite{10} have
pointed out the equivalence of the LPS model with wetting
models. The standard disordered wetting model \cite{4,5} has
been thoroughly studied and its solution is still challenging.
In this letter we establish a precise connection between a Levy type wetting
model and the LPS model, so that a first order transition is possible
in the ordered case.
Our results are based on a representation of the disordered
LPS model as a random matrix model, making contact with a 
localization problem. This representation allows to obtain an
exact supersymmetric formulation of the model, although here
we discuss a less rigorous one, based on replicas. We obtain
a mean--field approximation, which confirms that the critical
exponents are unchanged if $1 < \alpha < 3/2$, but if
$3/2 < \alpha < 2$ the exponents are independent of $\alpha$.
However, if $2 < \alpha < 3$ the phase transition remains first--order,
thus indicating that in this case fluctuations play a crucial
role.

{\it The disordered case} - 
DNA denaturation is described in the LPS model by the 
presence or the absence of an H(Hydrogen)-bond at the 
lattice site $x$ in a chain
of length $N$. A weight $\lambda_x$ is associated to the 
H-bond at site $x$, whereas the absence of H-bonds in a
sequence of sites of length $j$ is weighted by the algebraic
factor $a(j) = c j^{-\alpha}$. In fact, this model is
equivalent to a wetting model \cite{9}. Indeed, let us consider
a random walk that is constrained to the upper half-plane,
with a probability $p(m)$ of making a jump of size $m$. We can
assign a weight $\lambda_x$ at any time $x$ the random walk
touches the horizontal axis. There is a probability $a(j)$
that during time $j$ the walk does not touch the horizontal axis.
For the usual random walk, it is well known that asymptotically
$a(j) = c j^{- 3/2}$. More generally, we can prove that the same
results holds if $\sum_m p(m) m^2 < \infty$. On the other hand,
if the walk is of Levy type, i.e. if asymptotically 
$p(m) = {\bar p} m^{-\sigma}$ with  $1< \sigma <3$, then one
finds asymptotically $a(j) = c j^{-\frac{\sigma}{\sigma -1}}$.
This bridges the Levy type random walk with the LPS through
the relation $\alpha = \frac{\sigma}{\sigma -1} \ge 3/2$.
In both cases the thermodynamics is determined by a partition
function $q(N)$, that is given by the sum over all possible,
properly weighted configurations. A crucial quantity in the
wetting problem is the
density of contacts with the horizontal axis, $\rho$, which
corresponds to the density of H-bonds in the LPS model.
The phase transition is characterized by the fact that the
order parameter $\rho$ vanishes beyond a critical value of
the control parameter.

In both models, the partition function $q(N)$ obeys the
recursion formula
\begin{equation}
\label{eq4} 
\lambda_N q(N) = \sum_{x=0}^{N-2} q(N-1-x) a(x) + \delta_{N,1} 
\end{equation}
where for physical reasons $\lambda_x > 0$ and
$a(0) = 1$ in the LPS model. The ordered case corresponds to
$\lambda_x = \lambda \,\,\, , \,\,\, \forall x$: it can be
easily solved by computing the Laplace transform of
$q(N)$. One can then deduce from it the pressure 
$p = \lim_{N \to \infty} \frac{1}{N} \log q(N)$ and
the density $\rho$. One finds that $\rho = 0$ if
$\lambda > \lambda_c = \sum_{x =1 }^{\infty} a(x)$.
If $\lambda < \lambda_c$, near the critical point
the density vanishes as $\rho = {\bar\rho} (\lambda_c - \lambda)^
{\frac{2-\alpha}{\alpha -1}}$ if $1 < \alpha <2$. 
On the other hand, for $\alpha > 2$ one finds $\rho(\lambda_c) > 0$,
so that the transition is first order with a jump in the
order parameter.  It is worth stressing that in the Levy type random walk 
one has a first order transition for $ \frac{3}{2} < \sigma < 2$.
In \cite{12} we have also computed the structure factor of the
LPS model. We found that a basic length scale $\xi$ is associated to the
critical behavior of the model,
\begin{equation}
\label{eq6}
\xi \sim 
\begin{cases} 
|\lambda_c -\lambda|^{{\frac{1}{\alpha -1}}}
& \text{if} \,\,\, 1 < \alpha < 2 \\
|\lambda_c -\lambda|^{-1}
& \text{if} \,\,\, 2 < \alpha < 3
\end{cases} 
\end{equation}
In the limit $N \to \infty$ we find that the structure factor, $S(k)$, 
has a scaling limit
\begin{equation}
\label{eq7} 
\lim_{k \to 0, \xi\to\infty} \frac{S(k)}{S(0)} = F_{\alpha} (k \xi)
\end{equation}
where $F_{\alpha}$ is a universal function that depends only on the
scaling exponent $\alpha$, thus showing that $\alpha$ is a universal
critical exponent associated to this class of models.

{\it The disordered case} - 
According to some recent contributions by Giacomin and
Toninelli \cite{9} the disordered version of the LPS model
is expected to exhibit a continuos transition over the
whole range of values of $\alpha$. In order to tackle this problem
we exploit another useful representation of the partition function
as a determinant of a random matrix.
Indeed, let us define a perturbed partition function,
$q(N,\epsilon)$, where we
replace in the recursion formula (\ref{eq4}) $a(x)$ by $(1+\epsilon)a(x)$.
This allows to obtain the average density $\rho(N)$ as the logarithmic
derivative of $q(N,\epsilon)$ with respect to $\epsilon$ in the limit
of vanishing $\epsilon$. Let sus ntroduce the $N\times N$ matrix

\begin{equation}
\label{eq8} 
C_N^{\epsilon}(x,y)  = (1 + \epsilon) a(x-y) - \lambda_y \delta_{y,x+1},
\end{equation}
where $(x,y) \in \left[ 1,N \right]$ and
we have assumed that $a(x) = 0$ if $x \leq -1$. It can be
shown that the partition function can be rewritten as
\begin{equation}
\label{eq9} 
q(N,\epsilon) = \prod_{x = 1}^N \lambda_x^{-1} \det C_N^{\epsilon}
\end{equation}
Then the pressure and the density at finite volume can be expressed 
as follows
\begin{align}
\label{eq10}
& P_N = \frac{1}{N} \langle \ln q(N,0) \rangle \\
& \rho_N = \frac{1}{N} \frac{\mathrm d}{{\mathrm d} \epsilon} \langle \ln q(N,\epsilon) \rangle |_{\epsilon =0}
\end{align} 
where the symbol $\langle \bullet \rangle$ indicates the average over 
disorder. Let us observe that if $\lambda_x > 0 \,\,\, , \forall \,\,\, x$
then $ \det C_N^{\epsilon}$ is also positive.

On the other hand, one can extend these formulae to the case where
$\lambda_x$ is a gaussian variable with a given average $\langle \lambda
\rangle = \bar\lambda$ and with a given variance 
$ v = \langle \lambda_x^2 \rangle - \bar\lambda^2$.
For this purpose, we use the equality
\begin{equation}
\label{eq11} 
\det C_N^{\epsilon} = \left[(\det C_N^{\epsilon})^t
C_N^{\epsilon} \right]^{\frac{1}{2}}
\end{equation}
In this way the model can be generalized to gaussian disorder by
redefining the finite volume pressure and density as follows
\begin{align}
\label{eq12}
& P_N^{\epsilon} = {\hat P}_N^{\epsilon} - 
\frac{1}{2N} \sum_x \langle \ln \lambda_x^2 \rangle \\
& \rho_N = \frac{1}{N} \frac{\mathrm d}{{\mathrm d} \epsilon} \langle \ln q(N,\epsilon) \rangle |_{\epsilon =0}
\end{align} 
where
\begin{equation}
\label{eq13} 
{\hat P}_N^{\epsilon} =  {\frac{1}{2N}} \langle
\ln \det (C_N^{\epsilon})^t C_N^{\epsilon} \rangle
\end{equation}
We want to point out that the quantity $H = 
(C_N^{\epsilon})^t C_N^{\epsilon}$ can be considered as a random Hamiltonian.
Solving the problem of the thermodynamics of the disordered LPS model
amounts to compute the density of states of this random Hamiltonian.
In this context, the denaturation transition we are looking for 
can be interpreted as a passage from a phase of localized eigenstates ($\rho > 0$)
to a phase of extended eigenstates ($\rho =0$). This is a priori surprising
because in this case the properties of the density of states reflects the
localization transition.
amounts
to a localization transition from localized to extended eigenstates,
corresponding to $\rho > 0$ to $\rho =0$, respectively.
Usually, in this class of random Hamiltonian models the average over
disorder can be performed by two main techniques: the 
supersymmetric method and the replica trick. The former method is
certainly more rigorous and we shall present it in a future publication.
In what follows we
shall use the method of replicas, because it yields a more
transparent interpretation of the results. In practice, we aim at computing
the quantity
\begin{equation}
\label{eq14} 
Z_N^{\epsilon} = \langle (\det (C_N^{\epsilon})^t C_N^{\epsilon})^n \rangle
\end{equation}
so that we can obtain an explicit expression of the normalized pressure 
through the relation
\begin{equation}
\label{eq15} 
{\hat P}_N^{\epsilon} = \lim_{n \to 0} \frac{1}{2n} \ln Z_N^{\epsilon} 
\end{equation}
On the other hand, $ Z_N^{\epsilon}$ can be rewritten as follows
\begin{equation}
\label{eq16} 
Z_N^{\epsilon} = \int {\mathcal{D}} \psi \langle 
\exp{( \sum_{x,y} \left[\bar\psi_y, \psi_x\right] C_N^{\epsilon} (x,y))} \rangle
\end{equation}
where
\begin{equation}
\label{eq17} 
\left[\bar\psi_y, \psi_x\right] = \sum_{m=1}^{2n} \bar\psi_y(m) \psi_x(m)
\end{equation}
where $\{ \bar\psi_y(m)$ and $\psi_x(m) \}$ are grassman variables.
Now we can consider $\lambda_x$ as gaussian random variables or 
perform the weak-disorder limit. One obtains the expression
\begin{eqnarray}
\label{eq18} 
&Z_N^{\epsilon} = \int {\mathcal{D}} \psi 
\exp{( \sum_{x,y} \left[\bar\psi_y, \psi_x\right] D_N^{\epsilon}(x,y) )}
\nonumber
\\
&\times \exp{ (\frac{v}{2}\sum_x 
\left[\bar\psi_x, \psi_{x+1}\right]^2)}
\end{eqnarray}
where
\begin{equation}
\label{eq19} 
D_N^{\epsilon}(x,y) = (1 + \epsilon) a(x-y) - \bar\lambda  \delta_{y,x+1}
\end{equation}
Making use of the identity 
\begin{eqnarray}
\label{eq20} 
&\exp{( - \frac{v}{2} \mathrm{tr} H_x^2 )}= \\ 
&\frac{1}{d} \int {\mathrm d}Q_x \exp{( - 
\mathrm{tr} Q_x^2 - i \sqrt{v} \mathrm{tr} Q_x H_x)} \,\, ,
\nonumber
\end{eqnarray}
where $H_x$ and $Q_x$ are symmetric $(m \times m)$ matrices and $d$
is a suitable normalization constant, we can linearize the term
$\mathrm{tr} H_x^2 = - \frac{v}{2} \sum_x \left[\bar\psi_x, \psi_{x+1}\right]^2$ in 
Eq. (\ref{eq18}) and thus eliminate any dependence on the grassman
variables. We finally obtain:
\begin{eqnarray}
\label{eq21} 
&Z_N^{\epsilon} = 
\frac{1}{d^{N-1}} \int \prod_{x=1}^{N-1} d Q_x \exp( - \sum_x
\mathrm{tr} Q_x^2 
\nonumber \\
&\det \left[D_N^{\epsilon} + 2 i \sqrt{v} K\right] )
\end{eqnarray}
Here $D_N^{\epsilon} (x,m;x',m') = \delta_{m,m'} D_N^{\epsilon} (x,x')$ and
$K_N(x,m;x',m') = \delta_{x',x+1} Q_x (m,m')$. Since we want to focus our attention
to the weakly disordered case, we can perform the mean--field theory of the
problem at hand that can be obtained by the following steps:

\begin{itemize}
\item (s1) one has to first shift the random matrix:  $Q_x \to Q_x + i \sqrt{v} q_x 
\mathbb{I} $

\item (s2) the determinant in Eq.(\ref{eq21})
can be rewritten as $\det {\hat D}_N^{\epsilon}  \times \exp ({ \mathrm{tr} (\ln(\mathbb{I} + 2i \sqrt{v} 
({\hat D}_N^{\epsilon})^{-1} K)}) $ where 
$ {\hat D}_N^{\epsilon} (x,m; x',m') = D_N^{\epsilon} (x, x') \delta_{m,m'} -2 v q_x
\delta_{x',x+1} \delta_{m,m'}$

\item
(s3)  one can finally expand the logarithm to first order in $v$ and choose
$q_x$ in such a way that any linear term in $Q_x$ disappears from the
expression of $ Z_N^{\epsilon} $ (see Eq.(\ref{eq21})).
\end{itemize}

One can decompose the partition function into one part corresponding
to the mean--field contribution ($MF$) and another one corresponding
to the fluctuation one ($Fl$), namely
\begin{equation}
\label{eq22} 
Z_N^{\epsilon} = Z_N^{\epsilon \,\,\, MF} \,\, Z_N^{\epsilon\,\,\, Fl}
\end{equation}
where
\begin{equation}
\label{eq23a} 
Z_N^{\epsilon \,\,\, MF} = \det {\hat D}_N^{\epsilon} \exp 
\left[ 2 n v  \sum_x q_x^2 \right]
\end{equation}
and 
\begin{eqnarray}
\label{eq23b} 
&Z_N^{\epsilon \,\,\, Fl} = \frac{1}{d^{N-1}}
\int \prod_x dQ_x \exp( - \sum_{n=2}^{\infty} \sum_{\{ x_1, \cdots, x_n\}}
\nonumber \\ 
&s_n (x_1, \cdots, x_n) \mathrm{tr} Q_{x_1}\cdots Q_{x_n})
\end{eqnarray}
with
\begin{eqnarray}
\label{eq23c} 
&s_2(x_1,x_2)= \delta_{x_1,x_2} - 2v ({\hat D}_N^{\epsilon})^{-1}(x_1,x_2)
({\hat D}_N^{\epsilon})^{-1}(x_2,x_1) \nonumber \\
&s_n(x_1,\cdots ,x_n)= \frac{1}{n} (-2 i \sqrt{v})^n \prod_{j=1}^n 
({\hat D}_N^{\epsilon})^{-1}(x_j,x_{j+1})\nonumber \\
&
\end{eqnarray}
and the latter expression holds for $n \geq 3$.
Let us illustrate the mean--field results we obtain after performing the limit $N \to \infty$, in such
a way that $q_x \to q$ and $\lambda = \bar\lambda + 2 v q$. The resulting self--consistent
equation reads
\begin{equation}
\label{eq24} 
q = \int_0^{2 \pi} \frac{d \theta}{2 \pi} \left[ \sum_{j=0}^{\infty} a_j \exp{i \theta (j+1)} - \lambda \right]^{-1}
\end{equation}
and the corresponding density is
\begin{equation}
\label{eq25} 
\rho = -1 + \lambda q 
\end{equation}
Accordingly, we find that the critical point is shifted by disorder
\begin{equation}
\label{eq26} 
\bar\lambda_c = \lambda_c + (1 - \rho_c) \frac{2 v}{\lambda_c}
\,\,\,\, , \,\,\,\, (\rho_c = 0 \,\, {\mathrm{for}} \,\, \alpha < 2)
\end{equation}
and, consistently with the Harris criterion (see \cite{11}), we find
that the critical exponent is the same as in the ordered case for
$1 < \alpha < 3/2$, in formulae
\begin{equation}
\label{eq27} 
\rho = \bar\rho (\bar\lambda_c - \bar\lambda)^{\frac{2 - \alpha}{\alpha -1}}
\end{equation}
On the other hand, for $\frac{3}{2} < \alpha < 2$ we find
\begin{equation}
\label{eq28} 
\rho = \frac{ \lambda_c}{2 v}(\bar\lambda_c - \bar\lambda) \,\,\, ,
\end{equation}
i.e. the critical exponent is 1 all over this range
of values of $\alpha$. On the other hand the Mean-Field solution
still predicts a first--order transition for $2 < \alpha < 3$.
This indicates that higher order correction to the MF solution have
to be taken into account in order to confirm that the transition
becomes continuous also in this range in the presence of disorder.
We guess that an effective technique to tackle this problem should 
be based on a renormalization approach, directly applied to the 
partition function of the disordered model. Since we are dealing
with a 1D models we expect that a decimation procedure can be worked out.
This notwithstanding, the functional forms that appear in
the above equations seems to challenge the possibility of a fully
analytic, even if approximated, study.


{\it Conclusions.}
Our approach offers the possibility to compute exactly the
critical exponents if we take into account fluctuations in the
weak disorder limit. The case $\alpha > 2$ is particularly challenging,
since the rigorous work of \cite{2} shows that large deviations,
in the probabilistic sense, are responsible  for the continuous nature of 
the transition.

{\it Acknowledgments.} RL acknowledges financial support from the italian 
MIUR-PRIN project n. 20083C8XFZ.

\end{document}